\documentclass[12pt,a4paper]{article}
\begin{document}
\large

\newpage
\begin{center}
{\bf FAMILY STRUCTURE OF LEPTONS AND THEIR CURRENTS
OF AN AXIAL-VECTOR NATURE}
\end{center}
\vspace{0.5cm}
\begin{center}
{\bf Rasulkhozha S. Sharafiddinov}
\end{center}
\vspace{0.5cm}
\begin{center}
{\it Institute of Nuclear Physics, Uzbekistan Academy of Sciences,
Tashkent, 702132 Ulugbek, Uzbekistan}
\end{center}
\vspace{1cm}

All neutrinos have a non-zero mass and regardless of whether a neutrino
is of the Dirac or Majorana type, it can possess both anapole and electric
dipole moments. Between the corresponding form factors there appears a
connection, for example, for neutrino scattering on spinless nuclei. We
discuss a theory in which a mass consists of vector and axial-vector
components responsible for separation of leptonic current into the vector
and axial-vector parts of the same charge or dipole moment. Such a model
can explain the absence of truly neutral neutrino vector interactions and
the availability of an axial-vector structure of a Majorana mass. Thereby,
it relates the two neutrinos of different nature. We derive an equation
which relates the masses to a ratio of the anapole and electric dipole form
factors of any lepton and its neutrino as a consequence of their unification
in families of doublets and singlets. This testifies in favour of the existence
of the left (right) dileptons and paradileptons of the axial-vector currents.
Each of them answers to conservation of an axial-vector charge and any
lepton flavour. Therefore, an axial-vector mass, anapole and electric
dipole moment of the neutrino become proportional, respectively, to an
axial-vector mass, anapole and electric dipole moment of a particle
of the same family.

\newpage
\begin{center}
{\it 1. Introduction}
\end{center}

The nature has been created so that to any type of charged lepton corresponds
a kind of the neutrino \cite{1}. Such pairs can constitute the leptonic
families of the left-handed $SU(2)_{L}$-doublets as well as of the
right-handed $SU(2)_{R}$-singlets. This gives the right to define
their family structure in the united form \cite{2}:
\begin{equation}
\pmatrix{\nu_{e}\cr e^{-}}_{L},
(\nu_{e}, \, \, \, \, e^{-})_{R}, \, \, \, \,
\pmatrix{\nu_{\mu}\cr \mu^{-}}_{L},
(\nu_{\mu}, \, \, \, \, \mu^{-})_{R}, \, \, \, \,
\pmatrix{\nu_{\tau}\cr \tau^{-}}_{L},
(\nu_{\tau}, \, \, \, \, \tau^{-})_{R}, ...,
\label{1}
\end{equation}
\begin{equation}
\pmatrix{{\bar \nu_{e}}\cr e^{+}}_{R},
({\bar \nu_{e}}, \, \, \, \, e^{+})_{L}, \, \, \, \,
\pmatrix{{\bar \nu_{\mu}}\cr \mu^{+}}_{R},
({\bar \nu_{\mu}}, \, \, \, \, \mu^{+})_{L}, \, \, \, \,
\pmatrix{{\bar \nu_{\tau}}\cr \tau^{+}}_{R},
({\bar \nu_{\tau}}, \, \, \, \, \tau^{+})_{L}, ....
\label{2}
\end{equation}

Each family distinguishes from others by an individual flavour \cite{3,4}.
There exist, therefore, the three $(l=e,$ $\mu,$ $\tau)$ lepton flavours:
\begin{equation}
L_{l}=\left\{ {\begin{array}{l}
{+1\quad \; for\quad \; l^{-}_{L}, \, \, \, \, \, l^{-}_{R}, \, \, \, \,
\nu_{lL}, \, \, \, \, \nu_{lR},}\\
{-1\quad \; for\quad \; l^{+}_{R}, \, \, \, \, l^{+}_{L}, \, \, \, \,
{\bar \nu_{lR}}, \, \, \, \, {\bar \nu_{lL}},}\\
{\, \, \, \, 0\quad \; for\quad \; remaining \, \, \, \, particles.}\\
\end{array}}\right.
\label{3}
\end{equation}

Conservation of all types of lepton flavours
\begin{equation}
L_{l}={\it const},
\label{4}
\end{equation}
or full lepton number
\begin{equation}
L_{e}+L_{\mu}+L_{\tau}={\it const}
\label{5}
\end{equation}
is practically not excluded \cite{5}. Its legality follows from the fact
that both (\ref{4}) and (\ref{5}) become possible owing to a formation
of the united dileptons
\begin{equation}
(l^{-}_{L}, {\bar \nu_{lR}}), \, \, \, \, (l^{-}_{R}, {\bar \nu_{lL}}),
\label{6}
\end{equation}
\begin{equation}
(l^{+}_{R}, \nu_{lL}), \, \, \, \, (l^{+}_{L}, \nu_{lR})
\label{7}
\end{equation}
and paradileptons
\begin{equation}
\{(l^{-}_{L}, {\bar \nu_{lR}}),
(l^{+}_{R}, \nu_{lL})\}, \, \, \, \,
\{(l^{-}_{R}, {\bar \nu_{lL}}), (l^{+}_{L}, \nu_{lR})\}
\label{8}
\end{equation}
of the vector nature \cite{2}. For example, in the $\beta$-decays
\begin{equation}
n\rightarrow p^{\pm}e^{\mp}{\bar \nu_{e}}(\nu_{e}), \, \, \, \,
p^{\pm}\rightarrow ne^{\pm}\nu_{e}({\bar \nu_{e}}),
\label{9}
\end{equation}
\begin{equation}
\mu^{\mp}\rightarrow e^{\mp}{\bar \nu_{e}}(\nu_{e})
\nu_{\mu}({\bar \nu_{\mu}}), \, \, \, \,
\tau^{\mp}\rightarrow e^{\mp}{\bar \nu_{e}}(\nu_{e})
\nu_{\tau}({\bar \nu_{\tau}}),
\label{10}
\end{equation}
as well as in other phenomena with vector currents. Such systems can
also explain the conservation of summed charge.

According to the theory of unification of fermions \cite{2}, the mass
$m_{\nu_{l}},$ charge $e_{\nu_{l}}$ and vector moment $\mu_{\nu_{l}}$ of the
neutrino are proportional, respectively, to the mass $m_{l},$ charge $e_{l}$
and vector moment $\mu_{l}$ of a particle of the same leptonic family
\begin{equation}
m_{\nu_{e}}:m_{\nu_{\mu}}:m_{\nu_{\tau}}=
m_{e}:m_{\mu}:m_{\tau},
\label{11}
\end{equation}
\begin{equation}
e_{\nu_{e}}:e_{\nu_{\mu}}:e_{\nu_{\tau}}=
e_{e}:e_{\mu}:e_{\tau},
\label{12}
\end{equation}
\begin{equation}
\mu_{\nu_{e}}:\mu_{\nu_{\mu}}:\mu_{\nu_{\tau}}=
\mu_{e}:\mu_{\mu}:\mu_{\tau}.
\label{13}
\end{equation}

However, it is known \cite{6,7} that the interaction of leptons with
virtual photon may be described by the vertex operator
\begin{equation}
\Gamma_{\mu}(p,p')=\Gamma_{\mu}^{V}(p,p')+\Gamma_{\mu}^{A}(p,p'),
\label{14}
\end{equation}
including both the vector $(V)$ and axial-vector $(A)$ parts:
\begin{equation}
\Gamma_{\mu}^{V}(p,p')=\overline{u}(p',s')[\gamma_{\mu}f_{1l}(q^{2})-
i\sigma_{\mu\lambda}q_{\lambda}f_{2l}(q^{2})]u(p,s),
\label{15}
\end{equation}
\begin{equation}
\Gamma_{\mu}^{A}(p,p')=\overline{u}(p',s')\gamma_{5}[\gamma_{\mu}
g_{1l}(q^{2})-i\sigma_{\mu\lambda}q_{\lambda}g_{2l}(q^{2})]u(p,s).
\label{16}
\end{equation}
Here $\sigma_{\mu\lambda}=[\gamma_{\mu},\gamma_{\lambda}]/2,$ $q=p-p',$ $p(s)$
and $p'(s')$ denote the four-momentum (helicity) of the neutrino before and
after the interaction. The functions $f_{1l}(q^{2})$ and $f_{2l}(q^{2})$
at $q^{2}=0$ give the full electric charge and vector dipole moment
of this particle \cite{8}
\begin{equation}
e_{l}=f_{1l}(0), \, \, \, \, \mu_{l}=f_{2l}(0),
\label{17}
\end{equation}
on which there exist earlier and comparatively new laboratory and
astrophysical restrictions \cite{9,10}. The value of $e_{l}$ for the
lepton (antilepton) and its neutrino (antineutrino) has the negative
(positive) sign.

The values of $g_{1l}(0)$ and $g_{2l}(0)$ define the static size of the
neutrino anapole \cite{11} and electric dipole moments:
\begin{equation}
{\it a_{l}}=\frac{1}{m_{l}}\left(\frac{g_{1l}(0)}
{f_{1l}(0)}\right)^{2}f_{2l}(0), \, \, \, \, d_{l}=g_{2l}(0).
\label{18}
\end{equation}

Of them ${\it a_{l}}$ can also be measured \cite{12}, and for $d_{l}$
some experimental and cosmological limits are known \cite{9,10}.

In the framework of the recent presentations about the nature of these
currents, $g_{1l}(q^{2})$ must be CP-symmetrical, but P-antisymmetrical
\cite{11}. The function $g_{2l}(q^{2})$ is C-even, but CP-odd \cite{13}.
The existence of $g_{1l}(q^{2})$ is incompatible with the gauge
invariance. In other words, the terms $g_{il}(q^{2})$ can exist only
in the case when the mirror symmetry is violated in the absence of gauge
invariance. This possibility is realized if a particle has a
self inertial mass \cite{14}.

A massive Dirac $(l=\nu_{D})$ neutrino can, therefore, possess each of
the discussed currents \cite{14,15,16}. Their structure at $e=|e|$ for the
light $(\nu_{D}=\nu_{e})$ neutrino in one-loop approximation has the form
\begin{equation}
f_{1\nu_{D}}(0)=
\frac{3eG_{F}m_{\nu_{D}}^{2}}{4\pi^{2}\sqrt{2}}, \, \, \, \,
f_{2\nu_{D}}(0)=
\frac{3eG_{F}m_{\nu_{D}}}{8\pi^{2}\sqrt{2}},
\label{19}
\end{equation}
\begin{equation}
g_{1\nu_{D}}(0)=
\frac{3eG_{F}m_{\nu_{D}}^{2}}{4\pi^{2}\sqrt{2}}, \, \ \, \,
g_{2\nu_{D}}(0)=\frac{3eG_{F}m_{\nu_{D}}}{4\pi^{2}\sqrt{2}}.
\label{20}
\end{equation}

From the general considerations, it follows that $f_{il}(q^{2})$ and
$g_{il}(q^{2})$ describe the neutrino vector and axial-vector form factors.
If this is so, the standard $SU(2)_{L}$$\otimes$$U(1)$ theory \cite{17} states
that the functions $f_{il}(q^{2})$ and $g_{il}(q^{2})$ are proportional to
$g_{V{l}}$ and $g_{A_{l}},$ namely, to the coupling constants of the neutrino
vector and axial-vector currents.

From such a point of view, each of (\ref{19}) and (\ref{20}) takes place
only in the framework of the $(V-A)$ model, in which
\begin{equation}
f_{il}(q^{2}) \rightarrow \frac{1}{g_{V_{l}}}f_{il}(q^{2})=f_{il}(q^{2}),
\label{21}
\end{equation}
\begin{equation}
g_{il}(q^{2}) \rightarrow \frac{1}{g_{A_{l}}}g_{il}(q^{2})=g_{il}(q^{2}),
\label{22}
\end{equation}
where and further $g_{V_{l}}=g_{A_{l}}=1.$ Therefore, the form factors
$f_{1l}(0)$ and $g_{1l}(0)$ must be identical \cite{18} parameters of
the $(V-A)$ theory of a massive neutrino \cite{19,20}. The account
of the latter leads us to the conclusion that
\begin{equation}
g_{1l}(0)-f_{1l}(0)=0.
\label{23}
\end{equation}

Insofar as a neutrino \cite{21} of the Majorana type $(l=\nu_{M})$
is concerned, we start from the fact \cite{22} that it has no vector
interaction $(g_{V_{\nu_{M}}}=0),$ and its axial-vector interaction is
stronger $(g_{A_{\nu_{M}}}=2g_{A_{\nu_{D}}})$ than of a Dirac fermion. At the
same time, with the aid of (\ref{22}), it is not difficult to see \cite{14,23}
that truly neutral neutrino $(\nu_{M}=\nu_{1})$ anapole and electric dipole
moments are, according to (\ref{20}), equal to
\begin{equation}
g_{1\nu_{M}}(0)=\frac{3eG_{F}m_{\nu_{M}}^{2}}{2\pi^{2}\sqrt{2}}, \, \, \, \,
g_{2\nu_{M}}(0)=\frac{3eG_{F}m_{\nu_{M}}}{2\pi^{2}\sqrt{2}}.
\label{24}
\end{equation}

Such a conclusion one can make by investigating the processes with nuclei
in which appear the most diverse relations between the properties of Dirac
and Majorana neutrinos. Their nature, as we shall see below, gives the
possibility to define the structure of masses of neutrinos of both types.

It is also relevant to include in the discussion the united dependence of
the axial-vector form factors of leptons of the same family. This allows
to elucidate the ideas of each of the laws of conservation of summed charge,
lepton flavours and full lepton number.

The above questions will be illuminated in this work by studying the behavior
of light neutrinos of a different nature and of electrons in the elastic
axial-vector scattering on a spinless nucleus as a consequence of the
availability of rest mass, anapole and electric dipole moments of elementary
particles and their longitudinal polarization.

\begin{center}
{\it 2. Relation of axial-vector currents
of neutrino and electron}
\end{center}

The amplitude of polarized massive Dirac and Majorana fermions scattering
by nuclei in the limit of one-photon exchange may be written as
$$M_{fi}^{E}=\frac{4\pi\alpha}{q^{2}}\overline{u}(p',s')\gamma_{5}
[\gamma_{\mu}g_{1l}(q^{2})-
i\sigma_{\mu\lambda}q_{\lambda}g_{2l}(q^{2})]\times$$
\begin{equation}
\times u(p,s)<f|J_{\mu}^{\gamma}(q)|i>.
\label{25}
\end{equation}
Here $l=e=e_{L,R},$ $\nu=\nu_{D}=\nu_{e}=\nu_{eL,R}$ or $\nu_{M}=\nu_{1}=
\nu_{1L,R},$ and $J_{\mu}^{\gamma}$ denotes the Coulomb current
of a nucleus \cite{24}.

According to (\ref{25}), the interaction cross section of longitudinal
neutrinos and electrons with the field of a spinless nucleus
has the following structure
$$\frac{d\sigma_{E}^{A_{l}}(\theta_{l},s,s')}{d\Omega}=
\frac{1}{2}\sigma^{l}_{o}\{(1+ss')g_{1l}^{2}+$$
\begin{equation}
+4m_{l}^{2}\eta_{l}^{-2}(1-ss')g_{2l}^{2}
tg^{2}\frac{\theta_{l}}{2}\}F^{2}_{E}(q^{2}),
\label{26}
\end{equation}
in which
$$\sigma_{o}^{l}=
\frac{\alpha^{2}cos^{2}\frac{\theta_{l}}{2}}{4E^{2}_{l}(1-\eta^{2}_{l})
sin^{4}\frac{\theta_{l}}{2}}, \, \, \, \, \eta_{l}=
\frac{m_{l}}{E_{l}},$$
$$E_{l}=\sqrt{p^{2}+m_{l}^{2}}, \, \, \, \,
F_{E}(q^{2})=ZF_{c}(q^{2}),$$
$$q^{2}=-4E_{l}^{2}(1-\eta_{l}^{2})sin^{2}\frac{\theta_{l}}{2}.$$
Here $\theta_{l}$ is the axial-vector scattering angle, $E_{l}$
is the particle energy, $F_{c}(q^{2})$ is the nucleus charge form factor
$(F_{c}(0)=1).$ The index $A_{l}$ indicates the absence of the neutrino
vector currents.

The terms $(1+ss')$ and $(1-ss')$ describe the axial-vector Coulomb
interactions of the left $(s=-1)$ - and right $(s=+1)$-handed neutrinos
leading to the scattering with $(s'=-s)$ or without $(s'=s)$ flip of their
spin. Taking this into account, we can write (\ref{26}) as
\begin{equation}
d\sigma_{E}^{A_{l}}(\theta_{l},s)=
d\sigma_{E}^{A_{l}}(\theta_{l},g_{1l},s)+
d\sigma_{E}^{A_{l}}(\theta_{l},g_{2l},s),
\label{27}
\end{equation}
\begin{equation}
\frac{d\sigma_{E}^{A_{l}}(\theta_{l},g_{1l},s)}{d\Omega}=
\frac{d\sigma_{E}^{A_{l}}(\theta_{l},g_{1l},s'=s)}{d\Omega}=
\sigma^{l}_{o}g_{1l}^{2}F_{E}^{2}(q^{2}),
\label{28}
\end{equation}
$$\frac{d\sigma_{E}^{A_{l}}(\theta_{l},g_{2l},s)}{d\Omega}=
\frac{d\sigma_{E}^{A_{l}}(\theta_{l},g_{2l},s'=-s)}{d\Omega}=$$
\begin{equation}
=4m_{l}^{2}\eta_{l}^{-2}\sigma^{l}_{o}g_{2l}^{2}
F_{E}^{2}(q^{2})tg^{2}\frac{\theta_{l}}{2}.
\label{29}
\end{equation}

Averaging over $s$ and summing over $s',$ one can reduce the cross
section (\ref{26}) to the following form
\begin{equation}
d\sigma_{E}^{A_{l}}(\theta_{l})=
d\sigma_{E}^{A_{l}}(\theta_{l},g_{1l})+
d\sigma_{E}^{A_{l}}(\theta_{l},g_{2l}),
\label{30}
\end{equation}
\begin{equation}
\frac{d\sigma_{E}^{A_{l}}(\theta,g_{1l})}{d\Omega}=
\sigma^{l}_{o}g_{1l}^{2}F_{E}^{2}(q^{2}),
\label{31}
\end{equation}
\begin{equation}
\frac{d\sigma_{E}^{A_{l}}(\theta,g_{2l})}{d\Omega}=
4m_{l}^{2}\eta_{l}^{-2}\sigma^{l}_{o}g_{2l}^{2}
F_{E}^{2}(q^{2})tg^{2}\frac{\theta_{l}}{2}.
\label{32}
\end{equation}

Thus, (\ref{27}) and (\ref{30}) would seem allow the conclusion that either
incoming neutrinos are strictly longitudinally polarized, or they possess
no polarization. But we can say that this is not quite so. The point is
that the spin structure of their interaction with the field of emission
can not be defined regardless of the medium properties where it originates.
In other words, among the incoming and scattered neutrinos one can find
both longitudinally polarized and unpolarized particles, each of which
has suffered a strong change in his spin nature. Under such circumstances,
a flux of outgoing neutrinos is a partially ordered set of the scattered
fermions.

So, we must recognize that (\ref{26}) constitutes the naturally united set
of the axial-vector cross sections:
$$d\sigma_{E}^{A_{l}}=
\{d\sigma_{E}^{A_{l}}(\theta_{l},g_{1l},s), \, \, \,\,
d\sigma_{E}^{A_{l}}(\theta_{l},g_{2l},s), \, \, \, \, $$
\begin{equation}
d\sigma_{E}^{A_{l}}(\theta_{l},g_{1l}), \, \, \, \,
d\sigma_{E}^{A_{l}}(\theta_{l},g_{2l})\}.
\label{33}
\end{equation}

Of course, this class with a partial order is a partially ordered
set \cite{25} if between some pairs of its elements
$d\sigma_{E}^{A_{l}}(\theta_{l},g_{1l},s)$
and $d\sigma_{E}^{A_{l}}(\theta_{l},g_{2l},s)$
there exists a relation
\begin{equation}
d\sigma_{E}^{A_{l}}(\theta_{l},g_{1l},s)\le
d\sigma_{E}^{A_{l}}(\theta_{l},g_{2l},s)
\label{34}
\end{equation}
such that
\begin{enumerate}
\item
$d\sigma_{E}^{A_{l}}(\theta_{l},g_{1l},s)\le
d\sigma_{E}^{A_{l}}(\theta_{l},g_{1l},s)$ (reflexivity),
\item
$d\sigma_{E}^{A_{l}}(\theta_{l},g_{1l},s)\le
d\sigma_{E}^{A_{l}}(\theta_{l},g_{2l},s)$ and
$d\sigma_{E}^{A_{l}}(\theta_{l},g_{2l},s)\le
d\sigma_{E}^{A_{l}}(\theta_{l},g_{1l},s)$ imply
$d\sigma_{E}^{A_{l}}(\theta_{l},g_{2l},s)=
d\sigma_{E}^{A_{l}}(\theta_{l},g_{1l},s)$ (antisymmetricality),
\item
$d\sigma_{E}^{A_{l}}(\theta_{l},g_{1l},s)\le
d\sigma_{E}^{A_{l}}(\theta_{l},g_{2l},s)$ and
$d\sigma_{E}^{A_{l}}(\theta_{l},g_{2l},s)\le
d\sigma_{E}^{A_{l}}(\theta_{l},g_{1l})$ say
$d\sigma_{E}^{A_{l}}(\theta_{l},g_{1l},s)\le
d\sigma_{E}^{A_{l}}(\theta_{l},g_{1l})$ (transitivity).
\end{enumerate}

From our earlier developments, we find \cite{14,23} that any massive neutrino,
regardless whether it is a Dirac or a Majorana particle, has simultaneously
both anapole and electric dipole. A non-zero value of one of elements of
class (\ref{33}) will, therefore, indicate to the existence of each of them.
This becomes possible owing to the structural dependence of all elements of
the set (\ref{33}) which establishes the relation (\ref{34}) so that it
satisfies one more the condition:
$d\sigma_{E}^{A_{l}}(\theta_{l},g_{1l},s)\le
d\sigma_{E}^{A_{l}}(\theta_{l},g_{2l},s)$ implies
$d\sigma_{E}^{A_{l}}(\theta_{l},g_{2l},s)\le
d\sigma_{E}^{A_{l}}(\theta_{l},g_{1l},s)$ (symmetry).
Thus, if (\ref{34}) is both symmetric and antisymmetric, it is more natural
to expect that it appears in (\ref{33}) as a relation of equality.

It should also be observed that the cross sections of the axial-vector
interactions of longitudinal and unpolarized leptons are not different
\begin{equation}
\frac{d\sigma_{E}^{A_{l}}(\theta_{l},s)}
{d\sigma_{E}^{A_{l}}(\theta_{l})}=1.
\label{35}
\end{equation}

At first sight, the latter gives the possibility to compare separately
the contribution of the currents $g_{1l}(q^{2})$ or $g_{2l}(q^{2})$
to the scattering cross sections of longitudinal polarized and unpolarized
particles. On the other hand, such a comparison of (\ref{27}) and (\ref{30})
shows that
\begin{equation}
\frac{d\sigma_{E}^{A_{l}}(\theta_{l},g_{il},s)}
{d\sigma_{E}^{A_{l}}(\theta_{l},g_{il})}=1,
\label{36}
\end{equation}
and consequently, the class (\ref{33}) having a partial order is of those
partially ordered sets, in which the transitivity of relation (\ref{34})
leads to an equality implied from its symmetricality and antisymmetricality.
At the same time, the set (\ref{33}) can possess each of these properties
even at any permutation of elements. This is of course intimately connected
with the character of their structure depending on nature of the
discussed types of interactions.

They show that between the contributions of $g_{1l}(q^{2})$ and
$g_{2l}(q^{2})$ to the cross sections both in (\ref{27}) and in (\ref{30})
the same identity takes place, for which the interratios of possible pairs of
elements of class (\ref{33}) establish four most diverse structural equations.

For elucidation of their ideas, it is desirable to apply to the two of them:
\begin{equation}
\frac{d\sigma_{E}^{A_{l}}(\theta_{l},g_{2l},s)}
{d\sigma_{E}^{A_{l}}(\theta_{l},g_{1l},s)}=1, \, \, \, \,
\frac{d\sigma_{E}^{A_{l}}(\theta_{l},g_{2l})}
{d\sigma_{E}^{A_{l}}(\theta_{l},g_{1l})}=1,
\label{37}
\end{equation}
or to the two remaining equalities.

Such a choice is based logically on the fact that the anapole $g_{1l}(q^{2})$
and electric dipole $g_{2l}(q^{2})$ correspond to the two forms of the same
regularity of an axial-vector nature of the same charged lepton which unites
all elements of the set (\ref{33}) in a unified whole.

Inserting the explicit values of $d\sigma_{E}^{A_{l}}$ in any
of Eqs. (\ref{37}), we find that
\begin{equation}
4m_{l}^{2}\frac{g_{2l}^{2}(q^{2})}{g_{1l}^{2}(q^{2})}
\frac{tg^{2}\frac{\theta_{l}}{2}}{\eta_{l}^{2}}=1.
\label{38}
\end{equation}

If choose a particle energy $E_{l}\gg m_{l},$ at which $\eta_{l}\rightarrow 0,$
then for the case $q^{2}\rightarrow 0$ when $\theta_{l}\rightarrow 0,$
the limit is
$$lim_{\eta_{l}\rightarrow 0,\theta_{l}\rightarrow 0}
\frac{tg^{2}\frac{\theta_{l}}{2}}{\eta_{l}^{2}}=\frac{1}{4},$$
because of which the solution (\ref{38}) takes the form \cite{23}
\begin{equation}
m_{l}\frac{g_{2l}(0)}{g_{1l}(0)}=\pm 1.
\label{39}
\end{equation}

Together with (\ref{38}), that implies that $g_{1l}(q^{2})$ and
$g_{2l}(q^{2})$ are the functions which have the different values and
dimensionality for the same value of the square of the four-dimensional
momentum transfer. According to (\ref{37}), this reflects the unified
nature of their interaction with the field of emission. As a consequence,
the scattered flux with a partial order consists of a partially ordered
set of outgoing fermions. They constitute the class (\ref{33}) so that
the relation (\ref{34}) in it be reflexive, symmetric, antisymmetric
and transitive.

It should be mentioned, however, that the vertex $\Gamma_{\mu}^{A}$
is defined, for example, in many articles as \cite{26}
\begin{equation}
\Gamma_{\mu}^{A}(p,p')=\overline{u}(p',s')\gamma_{5}[\gamma_{\mu}q^{2}
G_{1l}(q^{2})-i\sigma_{\mu\lambda}q_{\lambda}G_{2l}(q^{2})]u(p,s).
\label{40}
\end{equation}
Here $G_{1l}(0)$ gives the dimensional anapole: ${\it a_{l}}=G_{1l}(0).$
In other words, the difference in (\ref{16}) and (\ref{40}) is that
\begin{equation}
g_{1l}(q^{2})=q^{2}G_{1l}(q^{2}) \, \, \, \, and \, \, \, \,
g_{2l}(q^{2})=G_{2l}(q^{2}).
\label{41}
\end{equation}

With these conditions, (\ref{31}), (\ref{32}) and (\ref{36})
replace (\ref{37}) by
\begin{equation}
\frac{G_{2l}^{2}(q^{2})}{4m_{l}^{2}G_{1l}^{2}(q^{2})}
\frac{\eta_{l}^{2}tg^{2}\frac{\theta_{l}}{2}}
{(1-\eta_{l}^{2})^{2}sin^{4}\frac{\theta_{l}}{2}}=1.
\label{42}
\end{equation}

Taking into account that
$$lim_{\eta_{l}\rightarrow 0,\theta_{l}\rightarrow 0}
\frac{\eta_{l}^{2}tg^{2}\frac{\theta_{l}}{2}}
{(1-\eta_{l}^{2})^{2}sin^{4}\frac{\theta_{l}}{2}}=1,$$
it is not difficult to get from (\ref{42}) the following relation
\begin{equation}
\frac{G_{2l}(0)}{2m_{l}G_{1l}(0)}=\pm 1.
\label{43}
\end{equation}

Comparing (\ref{39}) and (\ref{43}), we find \cite{14} that
\begin{equation}
G_{1l}(0)=\frac{g_{1l}(0)}{2m_{l}^{2}}.
\label{44}
\end{equation}

In both definitions (\ref{18}) and (\ref{44}), as expected from
simple reasoning \cite{11}, the anapole can not change its own value,
so that there exists a relation among the parameters
\begin{equation}
\frac{1}{m_{l}}\left(\frac{g_{1l}(0)}
{f_{1l}(0)}\right)^{2}f_{2l}(0)=G_{1l}(0).
\label{45}
\end{equation}

The equations (\ref{23}), (\ref{44}) and (\ref{45})
is reduced to an equality \cite{23}
\begin{equation}
g_{1l}(0)-2m_{l}f_{2l}(0)=0.
\label{46}
\end{equation}

Its insertion in (\ref{39}) allows to conclude that
\begin{equation}
g_{2l}(0)-2f_{2l}(0)=0.
\label{47}
\end{equation}

Such a behavior of the neutrino vector and axial-vector moments, together
with (\ref{23}), testifies also about the availability of a latent structure
of lepton-photon vertex depending on the particle mass.

\begin{center}
{\it 3. Relation of massive neutrinos of different nature}
\end{center}

We see that in the case of truly neutral neutrinos $(l=\nu_{M}),$ it is
convenient to replace the individual dependence between the anapole and
electric dipole form factors (\ref{39}) by
\begin{equation}
g_{1\nu_{M}}(0)-m_{\nu_{M}}g_{2\nu_{M}}(0)=0.
\label{48}
\end{equation}

It reflects the availability of a unified structure of Dirac and Majorana
neutrino axial-vector currents. It is important to elucidate whether there
exists a connection of massive neutrinos of both types. The elastic
scattering processes
\begin{equation}
\nu(\nu_{D}, \nu_{M})+A(Z)\stackrel{\gamma} {\rightarrow}
\nu'(\nu_{D}', \nu_{M}')+A(Z),
\label{49}
\end{equation}
\begin{equation}
{\bar \nu}({\bar \nu_{D}}, {\bar \nu_{M}})+A(Z)\stackrel{\gamma}
{\rightarrow}{\bar \nu'}({\bar \nu_{D}'}, {\bar \nu_{M}'})+ A(Z)
\label{50}
\end{equation}
are particularly interesting, because the incoming fluxes include the
Dirac and Majorana neutrinos.

Of course, this presentation is not a standard one. The number of particles
and the structural phenomena originating in the processes (\ref{49}) or
(\ref{50}) coincide, as follows from considerations of symmetry. Such a
conformity takes place regardless whether the neutrino is a Dirac or a
Majorana fermion. Then it is possible, for example, to study the united
processes (\ref{49}) and (\ref{50}) in the presence of only a Dirac neutrino
anapole and electric dipole moment of a Majorana particle.

To investigate further, we make the following replacements
in Eq. (\ref{26}):
$$g_{1l}(q^{2}) \rightarrow \frac{1}{g_{A_{\nu_{D}}}}g_{1\nu_{D}}(q^{2}),$$
$$g_{2l}(q^{2}) \rightarrow \frac{1}{2g_{A_{\nu_{D}}}}g_{2\nu_{M}}(q^{2}),$$
$$m_{l}\rightarrow m_{\nu_{D}}\rightarrow m_{\nu_{M}}\rightarrow m_{\nu}.$$

Thus, only the part is obtained of a general picture of elastic scattering
in which it is additionally assumed that
\begin{equation}
2g_{1\nu_{D}}(0)-m_{\nu}g_{2\nu_{M}}(0)=0.
\label{51}
\end{equation}

The equations (\ref{48}) and (\ref{51}), together with the first of
Eqs. (\ref{20}) at $m_{\nu}=m_{\nu_{D}}=m_{\nu_{M}}$ lead us once again
to Eqs. (\ref{24}). The indications are in favour of a certain latent
regularity of the nature of mass.

\begin{center}
{\it 4. United dependence of the axial-vector form factors
of the neutrino and electron}
\end{center}

We now remark that as in (\ref{34}), any relation
\begin{equation}
d\sigma_{E}^{A_{l}}(\theta_{l},g_{1l},s)\sim
d\sigma_{E}^{A_{l}}(\theta_{l},g_{2l},s)
\label{52}
\end{equation}
defined between the elements $d\sigma_{E}^{A_{l}}(\theta_{l},g_{1l},s)$ and
$d\sigma_{E}^{A_{l}}(\theta_{l},g_{2l},s)$ of the set (\ref{33}) is a relation
of equivalence \cite{25} only in the case when it satisfies the conditions
\begin{enumerate}
\item $d\sigma_{E}^{A_{l}}(\theta_{l},g_{1l},s)\sim
d\sigma_{E}^{A_{l}}(\theta_{l},g_{1l},s)$ (reflexivity),
\item
$d\sigma_{E}^{A_{l}}(\theta_{l},g_{1l},s)\sim
d\sigma_{E}^{A_{l}}(\theta_{l},g_{2l},s)$ says
$d\sigma_{E}^{A_{l}}(\theta_{l},g_{2l},s)\sim
d\sigma_{E}^{A_{l}}(\theta_{l},g_{1l},s)$ (symmetricality),
\item
$d\sigma_{E}^{A_{l}}(\theta_{l},g_{1l},s)\sim
d\sigma_{E}^{A_{l}}(\theta_{l},g_{2l},s)$ and
$d\sigma_{E}^{A_{l}}(\theta_{l},g_{2l},s)\sim
d\sigma_{E}^{A_{l}}(\theta_{l},g_{1l})$ imply
$d\sigma_{E}^{A_{l}}(\theta_{l},g_{1l},s)\sim
d\sigma_{E}^{A_{l}}(\theta_{l},g_{1l})$ (transitivity).
\end{enumerate}

Such an equivalence relation separates the class (\ref{33}) to the possible
subsets so that they have no general elements. The set (\ref{33}) may be
symbolically presented as
\begin{equation}
d\sigma_{E}^{A_{l}}=
\{d\sigma_{E}^{A_{l}}(\theta_{l},s), \, \, \,\,
d\sigma_{E}^{A_{l}}(\theta_{l})\}.
\label{53}
\end{equation}

Its subclasses reflect just the fact that each of Eqs. (\ref{27}) and
(\ref{30}) constitutes a kind of a united set of cross sections:
\begin{equation}
d\sigma_{E}^{A_{l}}(\theta_{l},s)=
\{d\sigma_{E}^{A_{l}}(\theta_{l},g_{1l},s), \, \, \, \,
d\sigma_{E}^{A_{l}}(\theta_{l},g_{2l},s)\},
\label{54}
\end{equation}
\begin{equation}
d\sigma_{E}^{A_{l}}(\theta_{l})=
\{d\sigma_{E}^{A_{l}}(\theta_{l},g_{1l}), \, \, \, \,
d\sigma_{E}^{A_{l}}(\theta_{l},g_{2l})\}.
\label{55}
\end{equation}

These classes are, according to (\ref{35}), equal. Similar equality can exist
only in the case when the scattering corresponds either to a vector $(V_{l})$
or to an axial-vector $(A_{l})$ component of leptonic current.

Here it is relevant to note that any neutrino possesses simultaneously only
one of the currents, $V_{l}$ or $A_{l}.$ In this situation, the interference
between the two interactions of a different nature may serve as a certain
indication to the appearance of the unified system of the two massive
neutrinos of the most diverse currents \cite{27}.

It is not excluded, however, that regardless of the values of scattering
cross sections of polarized and unpolarized particles in the field of a
nucleus, their interratio for any charged lepton and its neutrino has the
same value. In our case, from (\ref{35}), we are led to the following relation:
\begin{equation}
\frac{d\sigma_{E}^{A_{\nu_{l}}}(\theta_{\nu_{l}},s)}
{d\sigma_{E}^{A_{\nu_{l}}}(\theta_{\nu_{l}})}=
\frac{d\sigma_{E}^{A_{l}}(\theta_{l},s)}
{d\sigma_{E}^{A_{l}}(\theta_{l})}.
\label{56}
\end{equation}

In conformity with ideas of Eqs. (\ref{36}),
\begin{equation}
\frac{d\sigma_{E}^{A_{\nu_{l}}}(\theta_{\nu_{l}},g_{i\nu_{l}},s)}
{d\sigma_{E}^{A_{\nu_{l}}}(\theta_{\nu_{l}},g_{i\nu_{l}})}=
\frac{d\sigma_{E}^{A_{l}}(\theta_{l},g_{il},s)}
{d\sigma_{E}^{A_{l}}(\theta_{l},g_{il})}
\label{57}
\end{equation}
expressing the circumstance that the sets (\ref{54}) and (\ref{55}) both for
$\nu_{l}$ and for $l$ consist of the same elements.

Thus, on the basis of (\ref{38}), we can relate with confidence the masses to
a ratio of the axial-vector currents of each charged lepton and its neutrino
if reflexivity, symmetricality and transitivity of an equivalence relation
(\ref{52}) hold regardless of a particle type, owing to which the interratio
of any pair of elements from subclasses (\ref{54}) and (\ref{55}) for
$\nu_{l}$ and $l$ coincides. That gives the right to establish four
more most diverse identities.

To show their features, one can use two of them:
\begin{equation}
\frac{d\sigma_{E}^{A_{\nu_{l}}}(\theta_{\nu_{l}},g_{2\nu_{l}},s)}
{d\sigma_{E}^{A_{\nu_{l}}}(\theta_{\nu_{l}},g_{1\nu_{l}},s)}=
\frac{d\sigma_{E}^{A_{l}}(\theta_{l},g_{2l},s)}
{d\sigma_{E}^{A_{l}}(\theta_{l},g_{1l},s)},
\label{58}
\end{equation}
\begin{equation}
\frac{d\sigma_{E}^{A_{\nu_{l}}}(\theta_{\nu_{l}},g_{2\nu_{l}})}
{d\sigma_{E}^{A_{\nu_{l}}}(\theta_{\nu_{l}},g_{1\nu_{l}})}=
\frac{d\sigma_{E}^{A_{l}}(\theta_{l},g_{2l})}
{d\sigma_{E}^{A_{l}}(\theta_{l},g_{1l})},
\label{59}
\end{equation}
or the two remaining relationships.

Because of (\ref{31}), (\ref{32}) and (\ref{36}), any of (\ref{58}) and
(\ref{59}) allows to derive the same equation:
\begin{equation}
m_{\nu_{l}}\frac{g_{2\nu_{l}}(0)}{g_{1\nu_{l}}(0)}=
\pm m_{l}\frac{g_{2l}(0)}{g_{1l}(0)}.
\label{60}
\end{equation}

Another possibility is that by inserting (\ref{41}) in (\ref{28}), (\ref{29}),
(\ref{31}) and (\ref{32}), one can also find from (\ref{58}) and
(\ref{59}) that
\begin{equation}
\frac{G_{2\nu_{l}}^{2}(q^{2})}{4m_{\nu_{l}}^{2}G_{1\nu_{l}}^{2}(q^{2})}
\frac{\eta_{\nu_{l}}^{2}tg^{2}\frac{\theta_{\nu_{l}}}{2}}
{(1-\eta_{\nu_{l}}^{2})^{2}sin^{4}\frac{\theta_{\nu_{l}}}{2}}=
\frac{g_{2l}^{2}(q^{2})}{4m_{l}^{2}g_{1l}^{2}(q^{2})}
\frac{\eta_{l}^{2}tg^{2}\frac{\theta_{l}}{2}}
{(1-\eta_{l}^{2})^{2}sin^{4}\frac{\theta_{l}}{2}}.
\label{61}
\end{equation}

Disclosure of the uncertainties that define (\ref{43}), leads us
from (\ref{61}) to the following result
\begin{equation}
\frac{G_{2\nu_{l}}(0)}{2m_{\nu_{l}}G_{1\nu_{l}}(0)}=
\pm \frac{G_{2l}(0)}{2m_{l}G_{1l}(0)}.
\label{62}
\end{equation}

Its comparison with (\ref{60}) at $G_{2l}(0)=g_{2l}(0)$ establishes (\ref{44})
and thereby confirms the fact that the existence both of an individual and of
the united connections of the axial-vector cross sections of charged lepton
and its neutrino scattering is, by itself, not excluded.

\begin{center}
{\it 5. Conclusion}
\end{center}

Our study of the behavior of massive neutrinos of Dirac and Majorana types
in a nucleus Coulomb field shows clearly that between the properties of
these particles there exist well defined relations. We have established an
individual \cite{23} and the united relations of the anapole and electric
dipole form factors of each charged lepton and its neutrino. Such
regularities, however, encounter many problems which reflect the
characteristic features of the latent nature of the inertial mass.

At the same time, it is clear that (\ref{23}), (\ref{44}), (\ref{46}) and
(\ref{47}) define the anapole and electric dipole moments of any
neutrino or lepton:
\begin{equation}
{\it a_{l}}=\frac{g_{1l}(0)}{2m_{l}^{2}}=\frac{e_{l}}{2m_{l}^{2}}, \, \, \, \,
d_{l}=g_{2l}(0)=\pm \frac{e_{l}}{m_{l}}.
\label{63}
\end{equation}

These values, together with (\ref{19}), (\ref{20}) and (\ref{24}) require
the elucidation of the ideas of each of the existing types of charges
and masses.

From the point of view of mass-charge duality, any of the electric $(E),$ weak
$(W)$ and strong $(S)$ charges testifies in favour of the existence of a kind
of inertial mass \cite{28}. The neutrino mass and charge are naturally united
in rest mass $m_{l}^{U}$ and charge $e_{l}^{U}$ equal to
all the mass and charge:
\begin{equation}
m_{l}=m_{l}^{U}=m_{l}^{E}+m_{l}^{W}+m_{l}^{S}+...,
\label{64}
\end{equation}
\begin{equation}
e_{l}=e_{l}^{U}=e_{l}^{E}+e_{l}^{W}+e_{l}^{S}+....
\label{65}
\end{equation}

In the framework of the standard electroweak theory, a Majorana $(l=\nu_{M})$
neutrino has no an electric mass $(m_{\nu_{M}}^{E}=0)$ nor a Coulomb charge
$(e_{\nu_{M}}^{E}=0).$ Usually, it is accepted that the axial-vector
terms of leptonic current (\ref{14}) appear owing to the weak interaction
\cite{26}. Therefore, it seems that form factors $g_{i\nu_{M}}(0)$ arise
at the expense of a truly neutral neutrino weak mass.

On the other hand, as known, a massive Dirac neutrino interaction with
a weak field of emission may be expressed by a neutral current \cite{17}
consisting of vector and axial-vector parts. According to the
mass-charge duality \cite{28}, this implies that not only the weak
charge $e_{l}^{W}$ of a particle but also its weak mass $m_{l}^{W}$
includes both vector and axial-vector components.

A given circumstance seems to indicate that the electromagnetic form factors
$f_{i\nu_{e}}(0)$ and $g_{i\nu_{e}}(0)$ appear, respectively, due to the
vector and axial-vector parts of the neutrino weak mass.

Furthermore, if it turns out that any dipole moment can exist only in
the presence of a kind of charge \cite{8}, from the point of view of each
lepton-photon vertex (\ref{15}) or (\ref{16}), it should be expected that
the functions $f_{il}(q^{2})$ and $g_{il}(q^{2})$ correspond in nature to
the vector and axial-vector components of the same well known Dirac $(i=1)$
or Pauli $(i=2)$ interaction.

It is clear from the above considerations that $f_{1l}(q^{2})$ and
$g_{1l}(q^{2})$ characterize the vector and axial-vector parts of the
electric charge, and $f_{2l}(q^{2})$ and $g_{2l}(q^{2})$ describe their
dipole moments. They explain the availability of a vector \cite{2} as well
as of an axial-vector component of the Coulomb mass.

Thus, it follows that each part $(K=E,$ $W,$ $S, ...)$ of the neutrino
mass contains the vector and axial-vector components. We can, therefore,
separate any of the united rest mass (\ref{64}) and charge (\ref{65}) into
the vector and an axial-vector parts:
\begin{equation}
m_{l}=m_{V_{l}}^{K}+m_{A_{l}}^{K},
\label{66}
\end{equation}
\begin{equation}
e_{l}=e_{V_{l}}^{K}+e_{A_{l}}^{K}.
\label{67}
\end{equation}

With regard to the question about truly neutral neutrinos, their mass
is strictly an axial-vector type. Therefore, a Majorana neutrino has
not a vector nature.

So, we have learned that (\ref{51}) relates an axial-vector part of Coulomb
mass to form factors of a Dirac particle anapole and a Majorana neutrino
electric dipole moment. Therefore, from the point of view of the suggested
theory of mass, each of the above established relations (\ref{23}), (\ref{46})
and (\ref{47}) between the vector and axial-vector currents of the two
neutrinos of the same leptonic families must be interpreted as an indication
to the existence of a kind of the unified system of the two left (right)-handed
neutrinos of a different nature from the same purely neutrino families
\cite{14}. They are of course the united parafermions
\begin{equation}
(\nu_{D}^{L}, {\bar \nu_{M}}^{R}), \, \, \, \,
(\nu_{D}^{R}, {\bar \nu_{M}}^{L}),
\label{68}
\end{equation}
\begin{equation}
({\bar \nu_{D}}^{R}, \nu_{M}^{L}), \, \, \, \,
({\bar \nu_{D}}^{L}, \nu_{M}^{R})
\label{69}
\end{equation}
which appear in the presence of the field of emission both of fermions
with $V_{l}$ currents and of fermions of $A_{l}$ currents. Their scattering
on nuclei is described also by the interference contribution
$f_{il}(q^{2})g_{il}(q^{2})$ of the interaction vector $f_{il}(q^{2})$
and axial-vector $g_{il}(q^{2})$ parts \cite{27,29}.

This convinces us here that the appearance of any self interference
term $g_{il}^{2}(q^{2})$ in the cross section (\ref{26}) can be explained
by the availability in all families of doublets or singlets of a hard
axial-vector connection between the two left (right)-handed particles
of a definite type. It has the crucial value for steadiness of individual
dileptons of an axial-vector nature
\begin{equation}
(l_{L}^{-}, l_{R}^{+}), \, \, \, \,
(l_{R}^{-}, l_{L}^{+}),
\label{70}
\end{equation}
\begin{equation}
(\nu_{D}^{L}, {\bar \nu_{D}}^{R}), \, \, \, \,
(\nu_{D}^{R}, {\bar \nu_{D}}^{L}),
\label{71}
\end{equation}
\begin{equation}
(\nu_{M}^{L}, {\bar \nu_{M}}^{R}), \, \, \, \,
(\nu_{M}^{R}, {\bar \nu_{M}}^{L}).
\label{72}
\end{equation}

One of sharply expressed features of the discussed types of interactions
is the equality of each of interratios of the possible pairs of elements
of subsets (\ref{54}) and (\ref{55}) in the processes with lepton and its
neutrino. This corresponds in nature to the definite flavour. In other words,
any of (\ref{58}) and (\ref{59}) is valid only for particles
of the same type of lepton.

As a consequence, the structural dependence (\ref{60}) is in favour of
coexistence of each charged lepton and its neutrino. They can, therefore,
constitute the naturally united families of the left-handed
$SU(2)_{L}$-doublets as well as of the right-handed
$SU(2)_{R}$-singlets \cite{2}.

To express the idea more clearly, it is desirable to present (\ref{60})
in the form
\begin{equation}
m_{\nu_{l}}g_{1l}(0)g_{2\nu_{l}}(0)-
m_{l}g_{1\nu_{l}}(0)g_{2l}(0)=0.
\label{73}
\end{equation}

Turning to the cross section (\ref{27}), we remark that the anapole $g_{1l}$
does not change the direction of the particle spin, while the electric dipole
moment $g_{2l}$ is responsible for its flip \cite{29}. The latter confirms
the fact that the same lepton cannot have simultaneously both left - and
right-handed helicities. Therefore, each interference term in (\ref{73})
implies the existence of a kind of connected system of the two types
of left (right)-handed leptons of the axial-vector currents from
the same families of doublets or singlets.

It is seen that the equations (\ref{60}) and (\ref{73}) relate the two
united dileptons of an axial-vector nature. Such systems can appear because
of conservation of lepton flavours, for example, in any of $\beta$-decays
(\ref{9}) and (\ref{10}) as well as in the scattering on spin-zero nuclei
\cite{27}, if among the incoming particles not only leptons but also
their neutrinos are present.

We have already mentioned that formation of the left (right) dileptons and
paradileptons in the processes with leptonic currents $f_{il}(q^{2})$ is
responsible for conservation of vector part of the electric charge.

According to these results, each of earlier experiments \cite{6,9,30}
about conservation of summed electric charge and any type of lepton
numbers in reactions (\ref{9}), (\ref{10}) and
\begin{equation}
\gamma e^{-}\rightarrow e^{-}\nu_{e}{\bar \nu_{e}}, \, \, \, \,
e^{-}e^{+}\rightarrow \nu_{e}{\bar \nu_{e}},
\label{74}
\end{equation}
\begin{equation}
\nu_{e}e^{-}\rightarrow \nu_{e}e^{-}, \, \, \, \,
{\bar \nu_{e}}e^{-}\rightarrow {\bar \nu_{e}}e^{-}
\label{75}
\end{equation}
may serve as the source of facts confirming the existence of dileptons
and paradileptons of a different nature.

By following the structure of the united relation (\ref{60}),
we have
\begin{equation}
\frac{m_{\nu_{l}}}{m_{l}}=
\frac{g_{1\nu_{l}}(0)}{g_{1l}(0)}
\frac{g_{2l}(0)}{g_{2\nu_{l}}(0)}.
\label{76}
\end{equation}

The latter together with the ideas of full lepton number conservation law
predicts the size of the neutrino axial-vector mass
\begin{equation}
m_{\nu_{e}}:m_{\nu_{\mu}}:m_{\nu_{\tau}}=
m_{e}:m_{\mu}:m_{\tau}.
\label{77}
\end{equation}

One can also find from (\ref{60}) with the help of (\ref{18})
and (\ref{44}) that
\begin{equation}
{\it a}_{\nu_{e}}:{\it a}_{\nu_{\mu}}:{\it a}_{\nu_{\tau}}=
{\it a}_{e}:{\it a}_{\mu}:{\it a}_{\tau},
\label{78}
\end{equation}
\begin{equation}
d_{\nu_{e}}:d_{\nu_{\mu}}:d_{\nu_{\tau}}=
d_{e}:d_{\mu}:d_{\tau}.
\label{79}
\end{equation}

Thus, unlike the earlier presentations on the families of leptons \cite{31,32},
the discussed theory of unification of fermions \cite{2} leads us to a
correspondence principle that the axial-vector mass, anapole and electric
dipole moment of the neutrino are proportional, respectively, to the
axial-vector mass, anapole and electric dipole moment of a particle
of the same family.

Finally, the experimental observation of the above noted regularities
in the nature of fermions appears to be possible by measuring the effects
of the nuclear charge field \cite{2,14}. Of course, to make such a subtle
measurement one should create devices with sufficiently high sensitivity.

\end{document}